\documentclass[twocolumn,pra,superscriptaddress,showpacs]{revtex4}
\usepackage{graphicx}
\usepackage{dcolumn}
\usepackage{amsmath}
\usepackage{bm}
\usepackage{color}

\newcommand{\Hetwos}{$2\,{}^3S_1$}
\newcommand{\Hetwop}[1][j]{$2\,{}^3P_#1$}

\begin{document}

\title{Photoassociation spectra and the validity of the dipole approximation for 
weakly bound dimers}
\author{Daniel G. Cocks}
\affiliation{School of Engineering and Physical Sciences,
James Cook University, Townsville, Australia 4811}
\author{Ian B. Whittingham}
\affiliation{School of Engineering and Physical Sciences,
James Cook University, Townsville, Australia 4811}

\date{\today}

\begin{abstract}
Photoassociation (PA) of ultracold metastable helium to the $2s2p$ 
manifold is theoretically investigated using a non-perturbative close-coupled
treatment in which the laser coupling is evaluated without assuming the dipole 
approximation. The results are compared with our previous study 
[Cocks and Whittingham, Phys.~Rev.~A \textbf{80}, 023417 (2009)] that 
makes use of the dipole approximation.  The approximation is found to strongly 
affect the PA spectra because the photoassociated levels are weakly bound, and a 
similar impact is predicted to occur in other systems of a weakly bound nature.  
The inclusion or not of the approximation does not affect the 
resonance positions or widths, however significant differences are observed in 
the background of the spectra and the maximum laser intensity at which 
resonances are discernable. Couplings not satisfying the dipole selection rule  
$|J-1| \leq J^\prime \leq |J+1|$ do not lead to observable resonances.
\end{abstract}

\maketitle

\section{Introduction}

Photoassociation (PA) is a powerful technique exploited by researchers to probe 
the fundamental interactions of ultracold quantum gases \cite{Thors87,Weiner99}.  
Of particular interest is PA in metastable helium where release of the large 
internal energy during collisions provides unique experimental investigation strategies.
We have recently analysed the PA process \cite{Cocks09} and obtained detailed 
information of the laser intensity dependence of the line shifts and widths of 
the resonance peaks in the PA spectra of ultracold metastable 
helium excited to the $J=1, 0_u^+$ rovibrational states in the \Hetwos~+~\Hetwop[0]
asymptote. Two variants of a full nonperturbative multichannel, close-coupled 
treatment were used to obtain the required scattering matrix $\bm{S}$ in the 
presence of the non-vanishing asymptotic radiative coupling, one based 
upon dressed states, and the other on a modified radiative coupling that vanishes
asymptotically. Although both methods gave nearly identical results for the line
shifts and widths, these peaks were superimposed on very significant backgrounds
that differed, especially at higher laser intensities. These significant backgrounds
are a direct consequence of the weakly bound nature of the excited levels in 
metastable helium and are not found in PA studies of other systems.

In common with all previous investigations of PA in ultracold gases, we assumed the 
dipole approximation in evaluating the matrix elements of the laser coupling. 
However, in light of the sensitivity of the background radiation loss to the
form of the radiative coupling at large distances, we revisit the validity of this 
approximation and its applicability to weakly bound levels in general.

The laser coupling term is given by
\begin{equation}
\label{cw-1}
\hat{H}_{\text{int}} = -(\frac{e}{m})\sum_{i=1,2} \hat{\bm{p}}_{i}\cdot 
\hat{\bm{A}}(\bm{r}_{i})
\end{equation}
where $\hat{\bm{p}}_{i}=-i\hbar \bm{\nabla}_{\bm{r}_{i}}$ and the vector potential 
at the position $\bm{r}_{i}$ of the ith electron is
\begin{equation}
\label{cw-2}
\hat{\bm{A}}(\bm{r}_{i}) = \sum_{\xi}[\bm{\mathcal{E}}_{\xi}(\bm{r}_{i})\;\hat{a}_{\xi} + 
\bm{\mathcal{E}}_{\xi}(\bm{r}_{i})^{*}\;\hat{a}^{\dag}_{\xi}].
\end{equation}
Here $\hat{a}^{\dag}_{\xi}\;(\hat{a}_{\xi})$ are the creation (annihilation) operators for
a photon of angular frequency $\omega_{\xi}$ and polarization $\bm{\epsilon}_{\xi}$ and
\begin{equation}
\label{cw-2b}
\bm{\mathcal{E}}_{\xi}(\bm{r}_{i})=\sqrt{\frac{\hbar}{2 \omega_{\xi}\epsilon_{0}
\mathcal{V}}} e^{i\bm{k}\cdot \bm{r}_{i}}\bm{\epsilon}_{\xi},
\end{equation}
where $\bm{k}$ is the wave-vector of the laser field, $e$ and $m$ are the electron 
charge and mass respectively and $\mathcal{V}$ is the normalization volume.  The 
dipole approximation makes the assumption that $\exp(i\bm{k} \cdot \bm{r}_i) 
\approx 1$.  If the laser coupling is only significant in the region of the 
excited rovibrational states this is a reasonable assumption to make, as the 
outer turning points of the ultra long-range $J=1, 0_u^+$ vibrational states of 
this investigation place an upper bound of $r_i < R/2 < 235$~$a_0$, where $R$ is 
the interatomic distance.  Since $k =1/3258.17 \, a_0^{-1}$ for the  
$2s\,{}^3S-2p\,{}^3P$ transition, this gives 
$0.92 < |\exp(i \bm{k} \cdot \bm{r}_{i})| < 1.08$. 

The multichannel calculation of our previous investigation, however, involves open 
channels of the metastable basis coupled to the excited state, and the $S$-matrix 
elements and loss profiles were obtained by matching the asymptotic forms of the 
open channels at distances $R>10^5$~$a_0$, much greater than the  
interatomic ranges of the vibrational states.  
The coupling to the excited state, even at very large ranges where the uncoupled 
bound wavefunction was negligible, still influenced the calculation due to the 
closeness of the bound levels to the excited state dissociation limit.  This
influence can be suppressed and the background to the PA spectra removed by 
artificially deepening the potential well.  The presence of this background 
introduced new features such as the elimination of resonance peaks by saturation 
of the laser intensity, and it is evident that the laser coupling must be 
treated more carefully in non-perturbative calculations of PA profiles.

The validity of the dipole approximation at large interatomic separations of 
$O(10^{5})\,a_{0}$ is questionable, at least when weakly bound levels are considered. 
In this paper we study the effects of the non-dipole contributions to the PA 
spectra in metastable helium by firstly deriving an expression for the matrix
elements of the laser coupling which fully includes the $\exp(i\bm{k} \cdot \bm{r}_i)$ 
factors and then re-do the multichannel calculations of \cite{Cocks09} 
using these matrix elements.  

It is worth noting that the exponential term that is ignored in the dipole 
approximation has the effect of introducing a momentum transfer from the 
absorbed photon to the molecule. For ultracold atoms this momentum transfer 
can be much larger than the initial momentum of the colliding atoms.

\section{Exact laser coupling}

\subsection{General form}
The close-coupled equations for the two colliding atoms in the applied laser field
involve the matrix elements of the laser coupling operator, $\hat{H}_\text{int}$ 
between basis states of the general form
\begin{equation}
\label{cw-3}
|\Psi_g \rangle \equiv R^{-1} G_g(R) |g \rangle |n,\omega,\bm{\epsilon}_\lambda 
\rangle \end{equation}
and
\begin{equation}
\label{cw-3a}
|\Psi_e \rangle \equiv R^{-1} G_e(R) |e \rangle |n-1,\omega,\bm{\epsilon}_\lambda \rangle.
\end{equation}
Here the molecular basis states $|a \rangle$, where $a=\{g,e\}$, are assumed to have 
no dependence upon the interatomic distance $R$ but may still depend upon 
the molecular orientation in the space-fixed frame, specified by the angles 
$(\theta, \phi )$, and the electronic coordinates $\bm{r}_{i}$. 
The laser field states are denoted by $|n,\omega,\bm{\epsilon}_\lambda\rangle$, 
representing $n$ photons of frequency $\omega$ and polarization 
$\bm\epsilon_\lambda$, where $\lambda = 0,\pm 1$ for $\pi,\sigma^\pm$ 
polarization.  We assume the laser is directed along the space-fixed $Oz$-axis.

We consider the matrix element
\begin{equation}
\label{cw-4a} V^\text{int}_{eg} \equiv \langle e | \langle 
n-1,\omega,\bm\epsilon_\lambda| \hat{H}_\text{int} |g\rangle 
|n,\omega,\bm\epsilon_\lambda\rangle .
\end{equation}
As only the term involving the annihilation operator $\hat{a}_\xi$ contributes, 
we have
\begin{align}
\label{cw-4}
V^\text{int}_{eg} &= -\frac{e}{m} \sqrt{\frac{n \hbar}{2\omega 
\epsilon_0 \mathcal{V}}} \langle e| \sum_i (\hat{\bm{p}}(\bm{r}_i) \cdot 
\bm\epsilon_\lambda )
e^{i\bm{k}\cdot\bm{r}_i} |g\rangle
\end{align}
where
\begin{equation}
\label{cw-5}
\langle n-1,\omega,\bm\epsilon_\lambda| \hat{a}_\xi 
|n,\omega,\bm\epsilon_\lambda \rangle = \delta_{\lambda\xi} \sqrt{n}
\end{equation}
has been used.

Evaluation of (\ref{cw-4}) requires the choice of a coordinate system.  
We specify the origin of the coordinate system to be the center of mass of the 
dimer (assumed to consist of identical nuclei), such that the position 
of nucleus $A$ is given by $-\bm{R}/2$ and nucleus $B$ by $\bm{R}/2$. The electron 
coordinates $\bm{r}_{i}$ can then be expressed as $\bm{r}_{i} = 
\hat{\eta}_{i} \bm{R} / 2 + \bm{r}^\prime_{i}$ where the operator $\hat{\eta}_i$ 
has eigenvalues of $\mp 1$ if electron $i$ is centered upon atom $A$ or $B$ 
respectively, and $\bm{r}^\prime_{i}$ is the position vector of electron $i$ with 
origin at its respective atom.  Similarly, the momentum operators become 
$\hat{\bm{p}}(r_{i}) = \hat{\eta}_{i} \hat{\bm{p}}(\bm{R}) / 2 + 
\hat{\bm{p}}(\bm{r}^\prime_{i})$.

The matrix element (\ref{cw-4}) then becomes
\begin{eqnarray}
\label{cw-6}
V^\text{int}_{eg} &= &
-\frac{e}{m}\sqrt{\frac{n\hbar}{2\omega\epsilon_0\mathcal{V}}} \sum_i \langle e| 
\left[ \hat{\eta}_i \frac{1}{2} \hat{\bm{p}}(\bm{R}) + 
\hat{\bm{p}}(\bm{r}^\prime_i) \right] \cdot \bm\epsilon_\lambda \nonumber \\
&& \times e^{\hat{\eta}_i i\bm{k}\cdot\bm{R}/2} e^{i\bm{k}\cdot\bm{r}^\prime_i} 
|g\rangle.
\end{eqnarray}
This expression can be simplified by separating the basis states into a 
rotational part which depends only upon $(\theta, \phi )$ and the electronic 
part that is independent of $\bm{R}$:  \begin{equation}
\label{cw-7}
|a \rangle = |\psi_{a}^\text{rot} \rangle |\psi_{a}^\text{el}\rangle.
\end{equation}
After action of the momentum operators, the matrix element $\langle e | \ldots 
|g \rangle $ inside the summation becomes
\begin{eqnarray}
\label{cw-8}
&&\langle \psi^\text{rot}_e| \frac{i\hat{\eta}^2_i \bm{k}\cdot 
\bm{\epsilon}_\lambda}{4} e^{\hat{\eta}_i i\bm{k}\cdot\bm{R}/2} \langle 
\psi^\text{el}_e| e^{i\bm{k}\cdot\bm{r}^\prime_i} |\psi^\text{el}_g \rangle 
|\psi^\text{rot}_g \rangle \nonumber \\
&+&\langle \psi^\text{rot}_e| \frac{\hat{\eta}_i}{2} e^{\hat{\eta}_i 
i\bm{k}\cdot\bm{R}/2} \langle \psi^\text{el}_e| e^{i\bm{k}\cdot\bm{r}_i^\prime} 
|\psi^\text{el}_g\rangle \hat{\bm{p}}(\bm{R}) \cdot \bm{\epsilon}_\lambda 
|\psi^\text{rot}_g\rangle \nonumber \\
&+&\langle \psi^\text{rot}_e| e^{\hat{\eta}_i  i\bm{k}\cdot\bm{R}/2} \langle 
\psi^\text{el}_e|i(\bm{k}\cdot\bm{\epsilon}_\lambda) 
e^{i\bm{k}\cdot\bm{r}^\prime_i} |\psi^\text{el}_g\rangle | \psi^\text{rot}_g 
\rangle \nonumber \\
&+&\langle \psi^\text{rot}_e| e^{\hat{\eta}_i i\bm{k}\cdot\bm{R}/2} \langle 
\psi^\text{el}_e| e^{i\bm{k}\cdot\bm{r}^\prime_i} \hat{\bm{p}}(\bm{r}^\prime_i) 
|\psi^\text{el}_g\rangle \cdot \bm{\epsilon}_\lambda | \psi^\text{rot}_g \rangle 
.
\end{eqnarray}
The first and third terms are zero since $\bm{k}\cdot\bm{\epsilon}_\lambda=0$.  
In the second and fourth terms we can assume $\exp(i\bm{k}\cdot\bm{r}^\prime_i) 
\approx 1$ as the inner product $\langle \psi^\text{el}_e|\ldots 
|\psi^\text{el}_g\rangle$ is non-negligible with respect to $\bm{r}_i^\prime$ 
only in the regions of the atomic electrons. As such, the second term is also 
zero because the ground and excited states are orthogonal.

To proceed, we use the expansion for a plane wave in terms of Legendre 
polynomials
\begin{eqnarray}
\label{cw-9}
e^{ \hat{\eta}_i i\bm{k}\cdot\bm{R}/2 } &= & \sum_p i^p (2p+1) j_p(kR/2) 
P_p(\hat{\eta}_i \cos \theta) \nonumber \\
&= & \sum_p i^p (2p+1) j_p(kR/2) (\hat{\eta}_i)^p D^{p*}_{00}
\end{eqnarray}
 where $D^j_{m^\prime m}\equiv D^j_{m^\prime m}(\phi,\theta,0)$ is a Wigner 
rotation matrix \cite{Brink68}. Recalling the equality 
$e \hat{\bm{p}}(\bm{r}^\prime_i) / m = i [\hat{H}_\text{mol}, \hat{\bm{d}}^i] / 
\hbar$, where $\hat{\bm{d}}^i = e \bm{r}_i$ is the atomic dipole operator for 
electron $i$, the matrix element becomes
\begin{eqnarray}
\label{cw-10}
V^\text{int}_{eg} &= & - \sqrt{\frac{n}{2\hbar\omega\epsilon_0\mathcal{V}}} 
\sum_{i,p}
\langle \psi^\text{rot}_e | i^{p+1} (2p+1) j_p(kR/2) D^{p*}_{00} \nonumber \\
&& {}\times \langle \psi^\text{el}_e | (\hat{\eta}_i)^p \left[ 
\hat{H}_\text{mol}, \hat{\bm{d}}^i \right] |\psi^\text{el}_g \rangle \cdot 
\bm\epsilon_\lambda | \psi^\text{rot}_g \rangle \nonumber \\
 &= & - \sqrt{\frac{n\hbar\omega}{2\epsilon_0\mathcal{V}}} \sum_{i,p}
\langle \psi^\text{rot}_e | i^{p+1} (2p+1) j_p(kR/2) D^{p*}_{00} \nonumber \\
&& {}\times \langle \psi^\text{el}_e | (\hat{\eta}_i)^p \hat{\bm{d}}^i 
|\psi^\text{el}_g \rangle \cdot \bm\epsilon_\lambda | \psi^\text{rot}_g \rangle  
.
\end{eqnarray}
In obtaining (\ref{cw-10}) we have assumed the difference between electronic energies 
$E_{e}-E_{g}$ is approximately $\hbar\omega$. 
 
The matrix element of $\hat{\bm{d}}^i$ is most easily evaluated in the molecular frame 
using spherical tensors. The expansion required is
\begin{eqnarray}
\label{cw-11}
\hat{\bm{d}}^i \cdot \bm\epsilon_\lambda &= & \sum_\mu (-1)^\mu 
(\bm{\epsilon}_\lambda)_{-\mu} \hat{d}^i_\mu \nonumber \\
&= & \sum_{\beta} (-1)^\lambda D^{1*}_{\lambda \beta} \hat{d}^i_\beta,
\end{eqnarray}
where the subscripts $\mu$ and $\beta$ denote the spherical tensor components in 
the space- and molecular-fixed frames respectively and 
$(\bm{\epsilon}_\lambda)_{-\mu}=\delta_{\lambda,\mu}$.  This expansion allows 
the complete separation of the rotational and electronic parts of the matrix 
element:
\begin{eqnarray}
\label{cw-12}
V^\text{int}_{eg} &=& -\sqrt{\frac{n\hbar\omega}{2\epsilon_0\mathcal{V}}} 
\sum_{p}i^{p+1} (2p+1) j_p(kR/2) \nonumber \\
& & {}\times \sum_{i \beta F} (-1)^\lambda  C^{1pF}_{\lambda 0 \lambda} 
C^{1pF}_{\beta 0 \beta} \langle \psi^\text{rot}_e |D^{F*}_{\lambda \beta} 
|\psi^\text{rot}_g \rangle \nonumber \\
& & {}\times \langle \psi^\text{el}_e | (\hat{\eta}_i)^p \hat{d}^i_\beta 
|\psi^\text{el}_g \rangle.
\end{eqnarray}
Here the two rotation matrices, $D^{1*}_{\lambda \beta} D^{p*}_{00}$, have been 
combined using standard angular momentum theory \cite{Brink68}.

\subsection{Explicit form}

The appropriate explicit basis states are the hybrid Hund case (c) states \cite{Cocks09}
\begin{eqnarray}
\label{cw-13}
|a\rangle &\equiv & |\gamma j_1 j_2 j \Omega_j J m_J w\rangle \nonumber \\
&\equiv & \sqrt{\frac{2J+1}{4\pi}} D^{J*}_{m_J 
\Omega_j}(\phi,\theta,0) |\gamma j_1 j_2 j \Omega_j w \rangle
\end{eqnarray}
where  $L_\alpha$, $S_\alpha$ and $\bm{j}_\alpha=\bm{L}_\alpha + \bm{S}_\alpha$ 
are the orbital, spin and total angular momenta of atom $\alpha$ and $\gamma = 
\{\gamma_1\gamma_2\}$ where $\gamma_\alpha = \{\bar{\gamma}_\alpha L_\alpha 
S_\alpha\}$ with all other necessary atomic quantum numbers specified by 
$\bar{\gamma}_\alpha$.  $\bm{j}=\bm{j}_1 + \bm{j}_2$ is the total electronic 
angular momentum and $\bm{J}=\bm{j}+\bm{l}$ is the total angular momentum of the 
atom including rotation $\bm{l}$.  The labels $m$ and $\Omega $ indicate 
projections along the space-fixed $Oz$ axis and intermolecular $OZ$ axis 
respectively, and $w$ represents the gerade $(w=0)$ and ungerade $(w=1)$ 
symmetry, respectively, of inversion through the center of charge of the 
molecule.

The evaluation of the dipole operator must be performed in the $LS$ basis under 
the correct symmetry considerations \cite{Burke99}, however the inclusion of 
$\hat{\eta_i}$ modifies this result. In similar fashion to \cite{Cocks09}, we 
use the $LS$ basis of $|\gamma LS \Omega_L \Omega_S w\rangle$, where $\bm{L} = 
\bm{L}_1 + \bm{L}_2$ and $\bm{S} = \bm{S}_1 + \bm{S}_2$, symmetrized according to
\begin{eqnarray}
\label{cw-15}
|\gamma LS \Omega_L \Omega_S w\rangle  & = & N_{\gamma_1 \gamma_2} \left[ 
|\gamma_1^A \gamma_2^B LS \Omega_L \Omega_S \rangle_- \right. \nonumber \\
& &  + \left. (-1)^{p_{LS}} |\gamma_2^A \gamma_1^B LS \Omega_L \Omega_S 
\rangle_- \right] .
\end{eqnarray}
Here the $A$ and $B$ indicate the atom to which the orbital configuration $\gamma_i$ 
belongs, $N_{\gamma_1\gamma_2}$ is a normalization factor and $p_{LS}=w_1 + 
w_2 + L_1 + L_2 - L + S_1 + S_2 - S + N + w$ \cite{Nikitin84}, where $w_i$ 
refers to the atomic inversion symmetry of $\gamma_i$. The subscript `$-$' 
indicates that the state has been antisymmetrized with respect to electron 
permutation:
\begin{eqnarray}
\label{cw-15c}
|\gamma_1^A \gamma_2^B L S \Omega_L \Omega_S\rangle_- & = & \frac{1}{\sqrt{2}} 
\left[|\gamma_1^A \gamma_2^B L S \Omega_L \Omega_S; \bm{r}_1, \bm{r}_2 \rangle  
\right. \nonumber  \\
&& - \left.  |\gamma_1^A \gamma_2^B L S \Omega_L \Omega_S; 
\bm{r}_2, \bm{r}_1 \rangle \right].
\end{eqnarray}
When $\Omega_L+\Omega_S=0$, the states $|\gamma_i^A \gamma_j^B L S \Omega_L 
\Omega_S \rangle_-$ must also be properly symmetrized with respect to 
$\hat{\sigma}_v$, the reflection operator of the electronic wave function 
through a plane containing the intermolecular axis, however this does not affect 
the action of $(\hat{\eta}_i)^p\hat{d}^i_\beta$ and will be ignored. For the 
$2s2s$ and $2s2p$ states, the symmetrization reduces to
\begin{equation}
\label{cw-15b}
|2s2s \rangle = \frac{1}{2} \left[1 + (-1)^{w-S}\right] |(2s)^A (2s)^B 0S 0 
\Omega_S \rangle_- 
\end{equation}
and
\begin{eqnarray}
\label{cw-15d}
|2s2p \rangle &= &\frac{1}{\sqrt{2}} \left[ |(2s)^A (2p)^B 1S \Omega_L \Omega_S 
\rangle_- \right. \nonumber \\
& & \left. {} + (-1)^{1-S+w} |(2p)^A (2s)^B 1S \Omega_L \Omega_S \rangle_- 
\right].
\end{eqnarray}
Note that the coefficients differ from \cite{Burke99} as we include spin in the 
symmetrization. From the above, the matrix element of $(\hat{\eta}_i)^p 
d^i_\beta$ in the $LS$ basis is
\begin{equation}
\label{cw-16}
\langle 2s2p| (\hat{\eta}_i)^p d^i_\beta|2s2s\rangle = 
\frac{d_\text{at}}{2\sqrt{2}} \left[ 1 + (-1)^{w^\prime + 1 + S + p} \right] 
\delta_{\beta\Omega_L^\prime}
\end{equation}
where $d_\text{at}$ is the atomic dipole moment, $w^\prime$ and 
$\Omega_L^\prime$ refer to the excited state and the condition $w-S=$ even is 
assumed.  The presence of $(-1)^p$ is a direct result of the inclusion of the 
$(\hat{\eta}_i)^p$ term.  

To convert to the original basis (\ref{cw-13}), the transformation (A.11) of 
\cite{Cocks09} is used.  The matrix element of 
$(\hat{\eta}_i)^p \hat{d}^i_\beta$ in the basis (\ref{cw-13}) reduces to
\begin{eqnarray}
\label{cw-16b}
\lefteqn{\langle \gamma^\prime j_1^\prime j_2^\prime j^\prime \Omega_j^\prime 
w^\prime| (\hat{\eta}_i)^p \hat{d}^i_\beta | \gamma j_1 j_2 j \Omega_j w 
\rangle} \qquad \nonumber  \\
&&= \frac{1}{2\sqrt{2}} d_\text{at} F^{j_1^\prime j_2^\prime j^\prime 
\Omega_j^\prime}_{1j\beta \Omega_j} \left[ 1 + (-1)^{w^\prime + 1 + j + p} 
\right]
\end{eqnarray}
where
\begin{eqnarray}
\label{cw-16c}
F^{j_1 j_2 j \Omega_j}_{LS\Omega_L\Omega_S} &= & \sqrt{(2S+1)(2L+1)(2j_1+1)(2j_2+1)} 
\nonumber \\
& & {} \times C^{LSj}_{m_L m_S m_j} \left\{
\begin{array}{ccc}
L_1 & L_2 & L \\
S_1 & S_2 & S \\
j_1 & j_2 & j
\end{array} \right\},
\end{eqnarray}
$\{\ldots\}$ is a Wigner 9-j coefficient, and dashed quantities refer to the 
$2s2p$ state and undashed quantities to the $2s2s$ state.

The matrix element of the rotational part of $V^\text{int}_{eg}$ is
\begin{eqnarray}
\label{cw-17}
\lefteqn{\langle \psi^\text{rot}_e | D^{F*}_{\lambda \beta} | \psi^\text{rot}_g 
\rangle}  \nonumber  \\
&&= \iint \! \mathrm{d}\Omega \frac{\sqrt{(2J+1)(2J^\prime+1)}}{4\pi} 
D^{J'}_{m_J^\prime \Omega_j^\prime} D^{F*}_{\lambda \beta} D^{J*}_{m_J 
\Omega_j}.
\end{eqnarray}
Using the properties of 
the rotation matrices, the integral can be evaluated \cite{Brink68} so that
\begin{eqnarray}
\label{cw-18}
\lefteqn{\langle \psi^\text{rot}_e| D^{F*}_{\lambda \beta} (\phi,\theta,0)
|\psi^\text{rot}_g\rangle} \qquad\qquad \nonumber  \\
&=& \sqrt{\frac{2J+1}{2J^\prime+1}} C^{JFJ^\prime}_{m_J \lambda m_J^\prime} 
C^{JFJ^\prime}_{\Omega_j \beta \Omega_j^\prime}.
\end{eqnarray}

The complete matrix element is therefore
\begin{eqnarray}
\label{cw-19}
V^\text{int}_{eg} &= & -\sqrt{\frac{I}{\epsilon_0 c}} 
\sqrt{\frac{2J+1}{2J^\prime+1}} \sum_{p} i^{p+1}(2p+1) j_p(kR/2) \nonumber \\
& & {} \times \sum_{F\beta} (-1)^\lambda  C^{1pF}_{\lambda 0 \lambda} 
C^{1pF}_{\beta 0 \beta} C^{JFJ^\prime}_{m_J \lambda m_J^\prime} 
C^{JFJ^\prime}_{\Omega_j \beta
\Omega_j^\prime}  \nonumber \\
& & {} \times F^{j_1^\prime j_2^\prime j^\prime \Omega_j^\prime}_{1j\beta 
\Omega_j} d_\text{at} \frac{1 + (-1)^{w^\prime+1+j+p}}{2}
\end{eqnarray}
where $I$ is the laser intensity.

The dipole approximation corresponds to setting $kR=0$, and therefore, 
since $j_{p}(0)=\delta_{p,0}$, to $p=0$ in equation (\ref{cw-19}) 
\footnote{A factor of $-i$ is missing from equation (B16) of \cite{Cocks09}}.
The $p=0$ term of (\ref{cw-19}) differs from the dipole approximation result by
the presence of $j_0(kR/2)$ which varies little over the region of the 
excited bound levels but does oscillate and decay significantly outside this region. 
Larger values of $p$ do not contribute strongly in the molecular 
region due to the small magnitude of $j_p(kR/2)$ for $p \neq 0$ but are significant 
in the asymptotic region.  Lastly, we note that the 
restriction $\Delta w = \pm 1$ arising from the dipole approximation is broken, 
as even values of $p$ contribute to couplings to gerade metastable states and 
odd values of $p$ couple to ungerade metastable states.

\section{Results}

We first perform the calculations using the same basis states for the 
$\sigma^-$ coupling set that are used in \cite{Cocks09}, i.e.  we calculate 
photoassociation from the gerade metastable $J=2$ channels to the $0_u^+$, $J=1$ 
adiabatic potential that asymptotes to $j=0$. Unlike the calculations of 
\cite{Cocks09}, the asymptotic radiation coupling is zero because the summation 
over $p$ only contributes for $0 \leq p \leq 4$ and over this finite sum the 
spherical Bessel functions $j_p(kR) \rightarrow 0$ for $kR \gg p$ and $kR \gg 
1$. This means that the dressed state formalism or the $R$-dependent coupling 
method need not be employed to solve the equations.  However, for consistency 
between the two calculations, the $R$-dependent tail-off is introduced exactly 
as it was used in \cite{Cocks09}. 

The photoassociation profiles represent the spontaneous loss from the excited state 
and are determined from the loss of unitarity of the $S$-matrix. The photon 
emission cross section for atoms colliding in the entrance channel $|\alpha \rangle $
is 
\begin{equation}
\label{cw-19a}
\sigma^\text{photon}_{\alpha} = \frac{\pi}{k_\alpha^2} \left( 1 
- \sum_{\alpha^\prime} |S_{\alpha^\prime \alpha}|^2 \right)
\end{equation}
where $\alpha$ and $\alpha^\prime$ enumerate the open channels and $k_\alpha$ is the 
wavenumber of channel $\alpha$. The photoassociation profile is then 
the average
\begin{equation}
\label{cw-19b}
\sigma^\text{photon} = \frac{1}{n_o} \sum_\alpha \sigma^\text{photon}_{\alpha} 
\end{equation}
where $n_o$ is the number of open channels.

\begin{figure}
\includegraphics[width=0.95\linewidth]{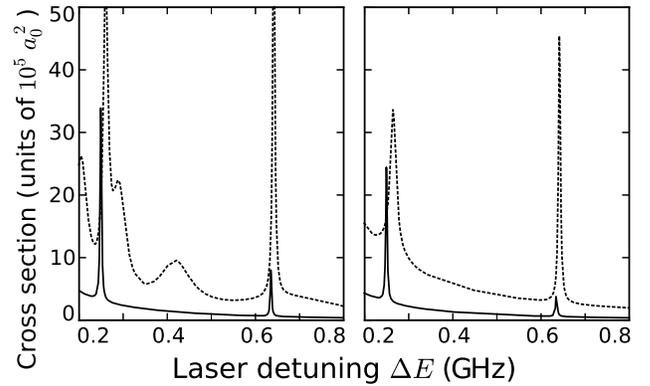}
\caption{\label{fig:l0l4_comp}
The photoassociation profiles for intensities of 64~mW/cm$^2$ (solid line) and 
640~mW/cm$^2$ (dotted line). The left hand plot includes only the $p=0$ term of the 
interaction coupling and hence only couplings to gerade metastable states. 
The right hand plot includes all terms of the $p$ summation and couplings to both 
gerade and ungerade metastable states.}
\end{figure}

Profiles calculated using only the $p=0$ contributions are shown in the left 
hand plot of Fig.~\ref{fig:l0l4_comp}.  The laser interaction term in this case 
differs from that in the dipole approximation by the presence of $j_0(kR/2)$. 
From the plot it is evident that, at large laser intensities, there are severe 
unphysical oscillations in the profiles.  Inclusion of all the terms in the 
$p$-summation that contribute ($p=0,2,4$) does not noticeably affect the 
profiles.

To mitigate this problem ungerade $S=1$ metastable states, normally uncoupled in  
the dipole approximation, are introduced as open channels.  
The $p=1,3$ terms of the laser coupling contribute to these channels and the 
total loss profiles for this combination of states are presented in the right 
hand plot of Fig.~\ref{fig:l0l4_comp}. These profiles exhibit more expected 
behavior and possess many similarities to the original PA profiles found 
using the dipole approximation. 

\begin{figure}
\includegraphics[width=0.95\linewidth]{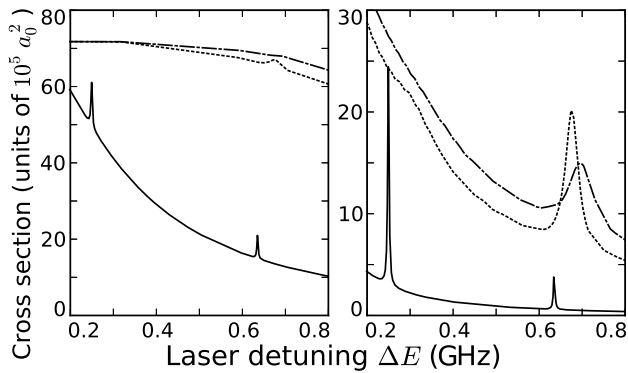}
\caption{\label{fig:l4DA_comp}
The photoassociation profiles for intensities of 64~mW/cm$^2$ (solid line), 
640~mW/cm$^2$ (dotted line) and 3.9~W/cm$^2$ (dash-dot line). The left hand plot 
shows profiles that result from the use of the dipole approximation whereas the 
right hand plot includes the full summation of the non-dipole interaction coupling.}
\end{figure}

Closer examination however (see Fig.~\ref{fig:l4DA_comp}) reveals that there is 
a large difference in behavior between the non-dipole and dipole approximation 
profiles, especially for high laser intensities.  At low intensities the 
non-dipole profiles show a reduction in the background loss but the resonance 
peaks have the same shift and width, albeit with an increased resonance 
strength.  At higher laser intensities, the resonance parameters still remain 
identical but the resonances in the non-dipole profiles remain discernable for 
larger intensities than those in the dipole approximation profiles.  In fact, 
the dipole approximation profiles demonstrate a saturation effect as the 
background overwhelms the resonance peaks. Instead, in the non-dipole profiles, 
we observe that the resonance peaks decrease in strength such that the overall loss 
in the resonance region is smaller at higher intensities.  This suggests possible 
optical suppression of photoassociative processes at high laser intensities.  
Why this occurs is unknown, although it is likely due to 
destructive interference between the ungerade and gerade metastable channels. 

The non-dipole coupling also introduces the possibility of more potential 
photoassociation channels, as the restriction $|J-1| \leq J^\prime \leq |J+1|$ 
is no longer enforced for summation terms $p \geq 1$. The likelihood that these 
couplings induce resonances in the excited state potentials is severely suppressed 
by the spherical Bessel functions as $j_p(kR/2) \ll 1$ for $p \neq 0$ and $kR \ll 
1$ (i.e. within the region of bound levels).   This prediction was tested by 
calculating profiles for photoassociation from the $J=2$ metastable levels to 
the $0_u^+$, $J=5$ excited levels and, as expected, no resonances were visible 
at any laser intensity.

In summary, the PA profiles presented here include all possible coupled metastable levels 
to the $0_u^+$, $J=1$ excited state and have been computed without assuming the 
dipole approximation for the laser coupling. The new calculations using the non-dipole 
coupling do not modify the resonance parameters that are tabulated in table II of \cite{Cocks09} 
to the accuracy of the calculations. The observed behavior of the background, 
and the influence of the dipole approximation, are not special to  
metastable helium but are a result of the weakly bound nature of the excited 
levels. It would be of much interest if a similar behavior were observed in 
photoassociation of another atomic species.

\appendix

\end{document}